\documentclass[aps,physrev,reprint,superscriptaddress]{revtex4-2}

\usepackage{graphicx}% Include figure files
\usepackage{amsmath}
\usepackage{siunitx}
\usepackage{dcolumn}% Align table columns on decimal point
\usepackage{bm}% bold math

\begin{document}

\title{\textbf{Exchange spin-wave propagation in Ga:YIG nanowaveguides}}

\author{Andrey A. Voronov}
\email{andrey.voronov@univie.ac.at}
\affiliation{Faculty of Physics, University of Vienna, Vienna, Austria}
\affiliation{Vienna Doctoral School of Physics, University of Vienna, Vienna, Austria}

\author{Khrystyna O. Levchenko}%
\affiliation{Faculty of Physics, University of Vienna, Vienna, Austria}

\author{Roman Verba}
\affiliation{V. G. Baryakhtar Institute of Magnetism of the NAS of Ukraine, Kyiv, Ukraine}

\author{Kristýna Davídková}
\affiliation{Faculty of Physics, University of Vienna, Vienna, Austria}
\affiliation{Vienna Doctoral School of Physics, University of Vienna, Vienna, Austria}

\author{Carsten~Dubs}
\affiliation{INNOVENT e.V. Technologieentwicklung, Jena, Germany}

\author{Michal Urbánek}
\affiliation{CEITEC BUT, Brno University of Technology, Brno, Czech Republic}

\author{Qi Wang}
\affiliation{School of Physics, Hubei Key Laboratory of Gravitation and Quantum Physics, Institute for Quantum Science and Engineering, Huazhong University of Science and Technology, Wuhan, China}

\author{Dieter Suess}
\affiliation{Faculty of Physics, University of Vienna, Vienna, Austria}

\author{Claas Abert}
\affiliation{Faculty of Physics, University of Vienna, Vienna, Austria}

\author{Andrii V. Chumak}
\affiliation{Faculty of Physics, University of Vienna, Vienna, Austria}

\date{\today}

\begin{abstract}
Spin-wave-based computing has emerged as a promising approach to overcome the fundamental limitations of CMOS technologies. However, the increasing demand for device miniaturization down to a \qty{100}{\nano\meter} scale presents significant challenges for long-distance spin-wave transport. Gallium-substituted yttrium iron garnet (Ga:YIG) offers a potential solution to these challenges due to its unique magnetic properties. The reduced saturation magnetization in Ga:YIG enables efficient excitation of exchange-dominated spin waves, which exhibit enhanced transport characteristics compared to dipolar-dominated modes in conventional materials. Here, we present the first comprehensive study combining experimental, analytical, and numerical investigations of spin-wave propagation in Ga:YIG waveguides down to \qty{145}{\nano\meter} width and \qty{73}{\nano\meter} thickness. Using micro-focused Brillouin light scattering spectroscopy, TetraX simulations, and analytical dispersion calculations, we demonstrate that Ga:YIG waveguides support spin waves with significantly higher group velocities up to \qty{600}{\meter \per \sec}. This value remains constant for structures with different widths, leading to longer spin-wave propagation lengths in nanowaveguides compared to non-substituted YIG. These results reveal that gallium substitution provides access to faster and longer-lived spin waves, opening new possibilities for implementing this material in nanoscale magnonic devices.
\end{abstract}

%\keywords{}

\maketitle

\section{Introduction}

Magnonics, the field that utilizes spin waves for information transport and processing, has emerged as a promising solution to overcome the fundamental limitations of complementary metal-oxide-semiconductor (CMOS) technologies~\cite{chumak2022advances, barman20212021}. Spin-wave-based computing offers several key advantages over conventional electronics, including scalability down to atomic dimensions, operation in the GHz-to-THz frequency range, and inherent nonlinearity for logic and beyond-Boolean operations~\cite{dieny2020opportunities, khitun2010magnonic, chumak2015magnon, nikolaev2024resonant}. Recent advances in inverse-design methods have further accelerated the development of magnonic logic devices by enabling the systematic design of functional elements with predetermined characteristics, opening new pathways for complex spin-wave circuits~\cite{voronov2025inverse, wang2021inverse, zenbaa2024experimental, zenbaa2025universal, papp2021nanoscale}.

The evolution of magnonic devices has progressed rapidly from initial millimeter-scale prototypes~\cite{schneider2008realization, fischer2017experimental, lee2008conceptual} to current nanoscale implementations~\cite{wang2018reconfigurable, wang2024nanoscale, khivintsev2022spin}. While recent demonstrations have shown successful operation of magnonic logic elements with sub-micrometer dimensions~\cite{wang2020magnonic}, the demands for further miniaturization continue to grow to meet the requirements of integrated circuit applications. This scaling down necessitates the development of increasingly narrow waveguides and thin-film magnetic structures. However, reducing waveguide dimensions typically leads to decreased spin-wave group velocities and shorter propagation lengths, even in low-damping materials such as yttrium iron garnet (YIG)~\cite{heinz2020propagation}. While switching to the Damon-Eshbach configuration can partially address these limitations, it introduces complications including challenging on-chip integration, complex multimode structures, and non-uniform field distributions across the waveguide cross-section~\cite{heinz2021long, bhaskar2020backward, levchenko20251d}. Alternative approaches using metallic ferromagnetic structures suffer from even higher magnetic damping, further limiting their practical applicability~\cite{demidov2011excitation, schwarze2013magnonic}.

A promising solution to these challenges lies in the utilization of gallium-substituted yttrium iron garnet (Ga:YIG)~\cite{dubs2025magnetically}, which provides access to long-wavelength exchange-dominated spin waves that remain detectable using conventional micro-focused Brillouin light scattering (\textmu BLS) techniques~\cite{breitbach2024nonlinear, bottcher2022fast}. The reduced saturation magnetization in Ga:YIG significantly increases the exchange length parameter, making exchange interactions dominant over dipolar interactions even at relatively long wavelengths. Such spin waves are easily excited and detected using conventional microwave and optical techniques, unlike more complex approaches involving exchange waves~\cite{wojewoda2023observing, wintz2016magnetic, yu2016approaching}. Crucially, the group velocity of exchange-dominated spin waves in Ga:YIG exhibits minimal dependence on waveguide width, making this material an ideal candidate for nanoscale magnonic applications~\cite{bottcher2022fast, carmiggelt2021electrical}.

In the present study, we experimentally investigate spin-wave propagation in Ga:YIG nanowaveguides with widths down to \qty{145}{\nano\meter}. Our results demonstrate that both propagation lengths and group velocities in these nanostructures are several times greater than those achieved in conventional YIG waveguides of comparable dimensions, despite the slight increase in Gilbert damping parameter that accompanies gallium substitution~\cite{heinz2020propagation}. Through a comprehensive approach combining \textmu BLS spectroscopy, numerical simulations, and analytical modeling in the backward-volume geometry, we establish the superior performance characteristics of Ga:YIG for nanoscale magnonic applications.

\section{Methods and materials}
\subsection{Ga:YIG nanowaveguides fabrication}

The nanowaveguides were fabricated from a \qty{73}{\nano\meter}-thick Ga:YIG film grown on a \qty{500}{\micro\meter}-thick (001)-oriented gadolinium gallium garnet (GGG) substrate via liquid phase epitaxy~\cite{dubs2025magnetically}. This growth technique ensures excellent crystalline quality and provides magnetic films with low Gilbert damping parameters~\cite{dubs2017sub}. Vibrating sample magnetometry (VSM) measurements of the Ga:YIG film revealed a saturation magnetization of $M_\mathrm{s} =\qty{22.76}{\kilo \ampere \per \meter}$, which is approximately one order of magnitude smaller than that of non-substituted YIG~\cite{dubs2020low}. This reduced magnetization yields a significantly increased exchange length $\lambda_\mathrm{ex} = (2 A_\mathrm{ex} / (\mu_0 M_\mathrm{s}^2))^{1/2}$, making exchange interactions dominant in governing spin-wave propagation in Ga:YIG~\cite{bottcher2022fast}.

An additional characteristic of the investigated material is the presence of stress-induced out-of-plane uniaxial anisotropy arising from the crystallographic lattice mismatch between the Ga:YIG film and the GGG substrate~\cite{zanjani2020predicting, dubs2025magnetically}. Ferromagnetic resonance (FMR) measurements determined the uniaxial anisotropy field strength as $\mu_0 H_\mathrm{an} = \qty[uncertainty-mode = separate]{73.6 \pm 0.6}{\milli\tesla}$, the gyromagnetic ratio $\gamma = \qty{178.8}{\radian \per \nano \second \per \tesla}$, and the Gilbert damping parameter $\alpha = \num{6.4e-4}$.

Nanowaveguides of varying widths were patterned on the Ga:YIG film using electron-beam lithography (\qty{30}{\kilo \electronvolt}, \qty{30}{\pico \ampere}) followed by argon ion-beam milling~\cite{levchenko20251d, wang2025fast}. For spin-wave excitation, coplanar waveguide (CPW) antennas~\cite{kalinikos1980excitation} were fabricated on top of the waveguides through electron-beam evaporation and subsequent lift-off processes. The example waveguide and antenna configuration, along with scanning electron microscopy (SEM) image of the nanofabricated structure, are shown in Fig.~\ref{fig:fig1}(b).

\subsection{Time-resolved \textmu BLS measurements}

\begin{figure}[b!]
    \centering
    \includegraphics[width=0.99\linewidth]{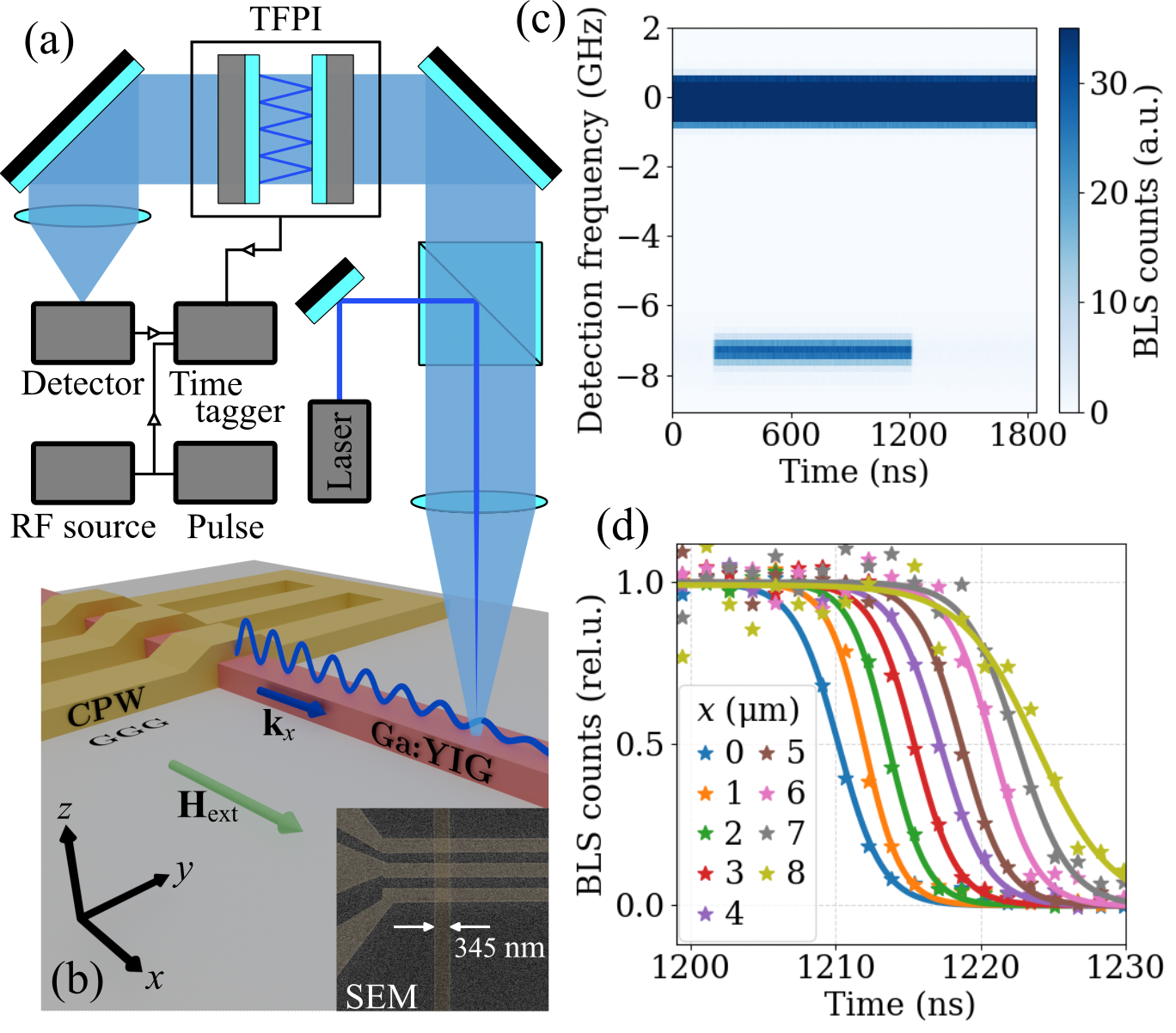}
    \caption{(a) Schematic representation of the \textmu BLS measurement system showing the laser beam path from the sample through the TFPI to the photodetector. The detection system is synchronized with the RF source and pulse generator to enable time-resolved measurements of spin-wave propagation. (b) Experimental configuration showing the Ga:YIG nanowaveguide orientation with respect to the external magnetic field $H_\mathrm{ext}$ applied in the backward-volume geometry, and the coplanar waveguide (CPW) antenna used for spin-wave excitation. The SEM image (inset) shows the fabricated nanowaveguide with width \qty{345}{\nano\meter}. (c) Time-resolved BLS spectrum showing the detection of a \qty{1000}{\nano\second} microwave pulse. The scattered magnon signal is detected at the spin-wave frequency, while the unscattered reference light appears at zero frequency. (d) Example time traces used for group velocity determination, measured over an \qty{8}{\micro\meter} propagation path from the antenna. The evolution of the pulse cut-off position at different spatial positions ($x$ = \qtyrange{0}{8}{\micro\meter}) demonstrates the spin-wave packet propagation along the nanowaveguide. A sigmoid function was used to approximate the data and determine the exact position of the falling edge of each pulse.}
    \label{fig:fig1}
\end{figure}

Spin-wave characterization was performed using \textmu BLS spectroscopy, a well-established technique for the investigation of spin-wave dynamics with high spatial and time resolution~\cite{sebastian2015micro, jorzick1999brillouin}. The experimental setup employs a continuous-wave single-frequency laser operating at \qty{457}{\nano\meter} wavelength, focused onto the sample through a microscope objective with high numerical aperture. The laser power was maintained at \qty[uncertainty-mode = separate]{3 \pm 0.2}{\milli\watt} on the sample surface to minimize thermal effects while ensuring sufficient signal intensity for reliable measurements~\cite{heinz2020propagation, wojewoda2023observing}. The inelastically scattered light is analyzed using a tandem Fabry-Pérot interferometer (TFPI), providing frequency resolution suitable for detecting spin-wave modes in the GHz range. The measured BLS intensity is directly proportional to the spin-wave intensity, enabling quantitative analysis of the magnonic properties.

To determine spin-wave group velocities, time-resolved \textmu BLS measurements were performed using the experimental setup shown in Fig.~\ref{fig:fig1}(a). The time-resolved technique relies on synchronized pulsed microwave excitation and photon detection to track spin-wave packet propagation. An RF source connected to a pulse generator produces microwave pulses that are applied to the CPW antenna to excite spin-wave packets in the nanowaveguide. The pulse generator simultaneously triggers the time-of-flight data acquisition system, establishing a precise temporal reference for the measurement~\cite{sebastian2015micro, buttner2000linear}. During the measurement, individual photons scattered from the propagating spin-wave packets are detected by the photodetector and their arrival times are recorded.

For the group velocity determination, microwave pulses with a duration of \qty{1000}{\nano\second} were applied, and the scattered signal corresponding to the spin-wave was then detected at the frequency of the CPW antenna excitation (\qty{6.9}{\giga\hertz} in this particular example), as shown in Fig.~\ref{fig:fig1}(c). This signal was normalized to the unscattered reference signal detected simultaneously at the same time. Subsequently, this measurement was performed at different positions along the waveguide over an \qty{8}{\micro\meter} path to determine the evolution of the falling edge of the pulse (Fig.~\ref{fig:fig1}(d)). To precisely determine the temporal position of the falling edge $t_0$, the acquired data at each spatial location was fitted with a sigmoid function:

\begin{equation}
    f(t) = \left[1+e^{a (t-t_0)}\right]^{-1},
\end{equation}

where the parameter $a$ controls the incline of the sigmoid function and is also being fitted. The group velocity was then determined from the falling edge time position evolution over the \qty{8}{\micro\meter} propagation distance by calculating the slope of the time-distance relationship.

Another important characteristic investigated using \textmu BLS was the spin-wave decay length in the nanowaveguides. To determine this parameter, the spin-wave intensity was measured along a \qty{20}{\micro\meter} path with a \qty{1}{\micro\meter} step size. The amplitude decay rate over distance was then analyzed and fitted using the exponential decay function:

\begin{equation}
    I(x) = I_1 \exp\left(-\frac{2x}{L_\mathrm{d}}\right) + I_0,
\end{equation}

where $I(x)$ is the measured BLS intensity at position $x$, $I_1$ is the initial spin-wave intensity close to the CPW antenna, $L_\mathrm{d}$ is the decay length, and $I_0$ is the background signal level. The factor of \num{2} in the exponent accounts for the fact that \textmu BLS measures the intensity of the spin waves, which is proportional to the square of the precession amplitude~\cite{heinz2020propagation}.

To ensure backward-volume geometry, an external magnetic field of \qty{255}{\milli\tesla} was applied parallel to the direction of spin-wave propagation. The experimental configuration on the sample is shown in Fig.~\ref{fig:fig1}(b).

\subsection{Numerical and analytical calculations}

To interpret the experimental results, the spin-wave dispersion relation $\omega (k_x)$ was calculated analytically in the backward-volume geometry using established theoretical framework~\cite{heinz2020propagation, wang2019spin, heinz2020temperature} while accounting for the uniaxial anisotropy present in Ga:YIG:

\begin{eqnarray}
    \label{eq:dispersion}
    \omega^2(k_x) =&& \left[\omega_H + \omega_M (F_y(k_x) + \lambda_\mathrm{ex}^2 k_x^2)\right]\nonumber \\
    &&\times \left[\omega_H - \omega_\mathrm{an} + \omega_M (F_z(k_x) + \lambda_\mathrm{ex}^2 k_x^2)\right],
\end{eqnarray}

where $\omega_M = \gamma \mu_0 M_\mathrm{s}$, $\omega_\mathrm{an} = \gamma \mu_0 H_\mathrm{an}$, $\omega_H = \gamma \mu_0 H_\mathrm{ext}$, $\lambda_\mathrm{ex}$ is the exchange length, $\mu_0$ is the vacuum permeability, and $F_y$, $F_z$ denote the dynamic demagnetization tensor components accounting for the waveguide geometry and shape anisotropy that it creates:

\begin{align}
    F_y &= \frac{1}{2 \pi} \int_{-\infty}^\infty \frac{k_y^2}{k^2} \frac{\sigma_k^2}{w} f(kh) \text{d}k_y \\
    F_z &= \frac{1}{2 \pi} \int_{-\infty}^\infty \frac{\sigma_k^2}{w} (1 - f(kh)) \text{d}k_y,
\end{align}

where $k^2 = k_x^2 + k_y^2$, $\sigma_k = w ~\text{sinc}(k_yw/2)$, $f(kh) = 1 - (1 - \exp(-kh))/(kh)$, $w$ is a width of the waveguide, and $h$ is its thickness. In the general case, dipolar interactions lead to spin pinning at the waveguide edges, which significantly complicates the dispersion relation. However, for narrow waveguides, the system can be considered in the unpinned state, substantially simplifying the analytical treatment~\cite{wang2019spin}. Therefore, Eq.~(\ref{eq:dispersion}) was implemented to analyze the fundamental spin-wave mode in the waveguide with width $w = \qty{145}{\nano\meter}$. 

Complementary fully-numerical simulations were performed using the finite-element micromagnetic modeling package TetraX~\cite{TetraX, korber2021finite}. These simulations combine the calculation of spin-wave dispersion and lifetime with the evaluation of CPW antenna absorption, providing a theoretical framework for comparison with experimental observations. The spin-wave group velocity was obtained as the derivative of the analytically and numerically obtained dispersion relation: $v_\mathrm{g} = \partial \omega / \partial k_x$. 

The material parameters used in the calculations were those determined from VSM and FMR measurements of the Ga:YIG film, with the exception of the saturation magnetization, which was adjusted to $M_\mathrm{s} = \qty{17.51}{\kilo \ampere \per \meter}$ to account for potential modifications during the nanofabrication process. The exchange stiffness $A_\mathrm{ex} = \qty{1.37}{\pico \joule \per \meter}$ was adopted from previous studies on similar Ga:YIG films~\cite{bottcher2022fast}.

The dispersion relation $\omega(k_x)$ obtained using Eq.~(\ref{eq:dispersion}) and the numerical simulations for the waveguide with width \qty{145}{\nano\meter} and thickness \qty{73}{\nano\meter} is presented in Fig.~\ref{fig:fig2}(a). Both theoretical and numerical approaches yield nearly identical results, demonstrating their reliability as robust analytical tools for investigating spin-wave behavior in nanoscale Ga:YIG waveguides.

\section{Results and discussion}

\begin{figure}[b]
    \centering
    \includegraphics[width=0.98\linewidth]{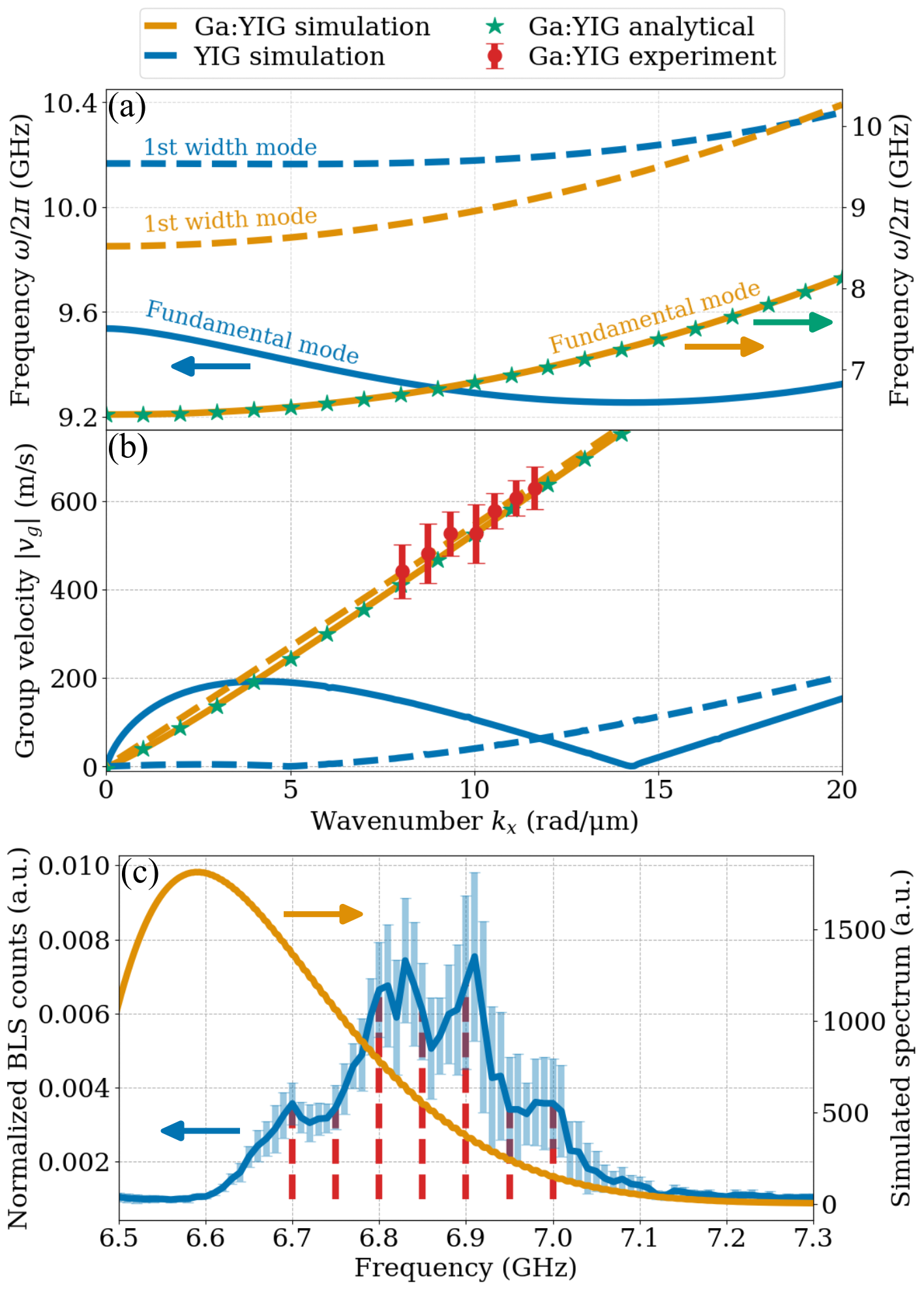}
    \caption{(a) Calculated dispersion relations $\omega(k_x)$ for non-substituted YIG and Ga:YIG obtained using the analytical formula (Eq.~(\ref{eq:dispersion})) and numerical simulations with TetraX. (b) Absolute values of group velocities $v_\mathrm{g}$ derived from the calculated dispersion relations, compared with experimentally determined values for Ga:YIG obtained from time-resolved \textmu BLS measurements. Solid lines in subfigures (a-b) indicate the fundamental spin-wave mode, while dashed lines refer to the simulations of the 1$^{\text{st}}$ width mode. (c) Experimentally measured frequency sweep (blue line) showing the spin-wave amplitude for excitation frequencies applied to the antenna, normalized to the amplitude of the elastically scattered signal. Additionally, the spin-wave spectrum was estimated using microwave absorption and lifetime simulation (orange line). Red dashed lines indicate the chosen excitation frequencies, which were subsequently used for the group velocity measurements. All the theoretical calculations and experimental data correspond to waveguide dimensions of \qtyproduct{145 x 73}{\nano \meter} (width × thickness).}
    \label{fig:fig2}
\end{figure}

Comparing the dispersion relations $\omega(k_x)$ for non-substituted YIG and Ga:YIG obtained from numerical simulations (Fig.~\ref{fig:fig2}(a)), a distinct difference in behavior is observed in the backward-volume geometry. Due to the significantly reduced saturation magnetization $M_\mathrm{s}$ in Ga:YIG and the correspondingly increased exchange length $\lambda_\mathrm{ex}$, the dispersion curve is almost proportional to $\omega_M\lambda_\mathrm{ex}^2 k_x^2$ not only for the fundamental spin-wave mode but also for other high-frequency width modes (dashed lines in Fig.~\ref{fig:fig2}). In contrast, the dispersion relation in non-substituted YIG is dominated by the dipolar interaction for low wavevectors, resulting in a characteristic negative slope.

This fundamental difference has a direct impact on the spin-wave group velocities $v_\mathrm{g}$ in these materials, as demonstrated in Fig.~\ref{fig:fig2}(b). While maximum values of $v_\mathrm{g}$ in the wavevector-range of interest in YIG reaches approximately \qty{200}{\meter \per \second} in waveguides with dimensions \qtyproduct{145 x 73}{\nano \meter}, the group velocity in Ga:YIG scales almost linearly with $k_x$. The scaling rate remains constant for higher width modes due to the significantly increased exchange interaction influence on the spin-wave dispersion. This behavior enables Ga:YIG to overcome the velocity limitations of non-substituted YIG and achieve spin waves with significantly higher group velocities at larger wave vectors.

To validate the numerical predictions, BLS measurements were conducted to determine the group velocities of spin waves in Ga:YIG nanowaveguides. Initially, the narrowest available waveguide with a width of \qty{145}{\nano\meter} was investigated. A frequency sweep measurement was performed to determine the accessible frequency range for spin-wave detection using BLS, as shown in Fig.~\ref{fig:fig2}(c). The detection was carried out at a distance of \qty{1}{\micro \meter} from the antenna, and the excitation power was set to \qty{-10}{\decibel}m to reduce the influence of non-linear phenomena on the spin-wave propagation. 

Additionally, the spin-wave spectra were estimated using TetraX simulations. It allows for the calculation of the spin-wave mode linewidth $\Gamma$ which is used to determine its wavevector-dependent lifetime $\tau$. The decay length $L_\mathrm{d}$ of the mode is then obtained from the lifetime $\tau$ and the spin-wave group velocity $v_\mathrm{g}$: $L_\mathrm{d} = v_\mathrm{g} \tau$. Combining this value with the CPW antenna absorption simulation provided by TetraX, one can estimate the spin-wave spectrum at the given distance from the antenna. The simulated in this way spin-wave spectrum demonstrates considerable overlap with the experimentally obtained frequency sweep (Fig.~\ref{fig:fig2}(c)), with the observed discrepancy attributed to the nanofabrication process, which typically leads to slight alterations in material parameters~\cite{heinz2020propagation}.

From the frequency sweep experiment, several frequencies were selected (indicated by red dashed lines in Fig.~\ref{fig:fig2}(c)) for subsequent time-resolved BLS measurements. The experimentally determined group velocities are presented in Fig.~\ref{fig:fig2}(b). The specific wavevectors corresponding to the excited frequencies were calculated from the numerically obtained dispersion curve for the investigated waveguide (Fig.~\ref{fig:fig2}(a)). The measured group velocities reproduce the linearly increasing behavior with wavevector predicted by the dispersion curve simulations, achieving group velocities up to \qty{600}{\meter \per \second}.

\begin{figure*}[t!]
    \centering
    \includegraphics[width=0.98\linewidth]{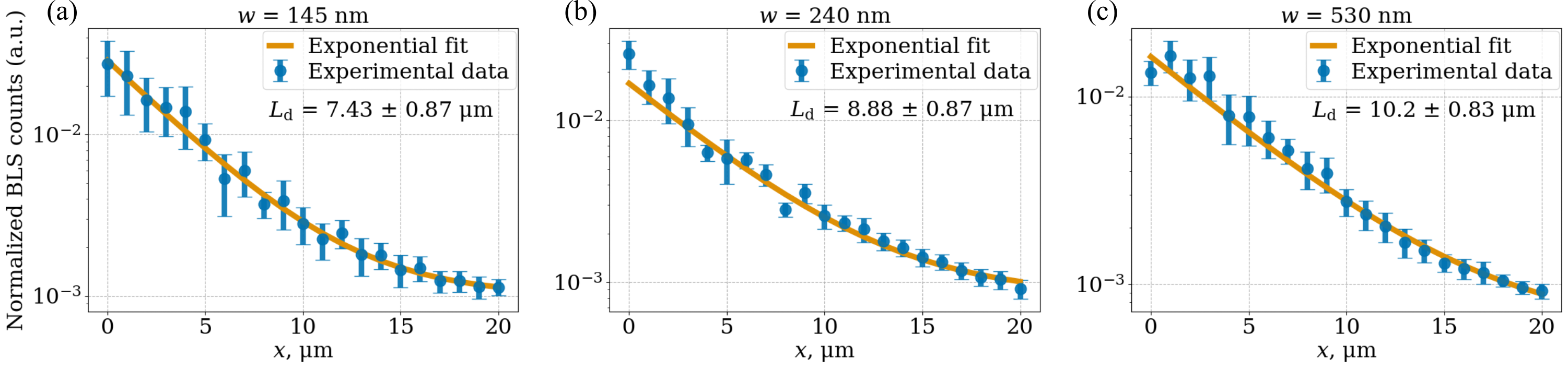}
    \caption{Spin-wave decay length measurements in Ga:YIG nanowaveguides with different widths: (a)~$w = \qty{145}{\nano\meter}$, (b)~$w = \qty{240}{\nano\meter}$, and (c)~$w = \qty{530}{\nano\meter}$. Blue circles represent experimental BLS data points with error bars, while orange lines show exponential fits used to determine the decay length $L_\mathrm{d}$. The BLS counts are plotted on a logarithmic scale as a function of propagation distance along the waveguide and normalized to the amplitude of the elastically scattered signal. The measurements are performed at the excitation frequency of \qty{6.95}{\giga \hertz}.}
    \label{fig:fig3}
\end{figure*}

Moreover, since Ga:YIG provides a platform for isotropic exchange-dominated spin waves, the internal demagnetization field distribution does not significantly affect the group velocity values when comparing nanowaveguides with different widths. Further \textmu BLS measurements revealed that in structures with widths of \qty{240}{\nano\meter} and \qty{530}{\nano\meter}, the group velocity values are remarkably similar: \qty[uncertainty-mode = separate]{529 \pm 56}{\meter \per \second} and \qty[uncertainty-mode = separate]{524 \pm 41}{\meter \per \second}, respectively. At the same time, the numerically predicted values for YIG nanostructures with same dimensions are \qty{106}{\meter \per \second} and \qty{97}{\meter \per \second}, respectively. The spin waves in these measurements were excited at the same wavevector of $k_x = \qty{11.25}{\radian \per \micro \meter}$ which roughly corresponds to the frequency of \qty{6.95}{\giga\hertz} in Ga:YIG nanowaveguides. The group velocity comparison between materials constitutes an almost 5-fold increase in Ga:YIG nanostructures across different widths. 

This difference in group velocity values between materials increases with decreasing nanowaveguide widths. Numerical simulations predict group velocity $v_\mathrm{g}$ in Ga:YIG of \qty{613}{\meter \per \second} and in YIG of \qty{70}{\meter \per \second} for a nanowaveguide with dimensions \qtyproduct{50 x 50}{\nano \meter}~\cite{heinz2020propagation}. These results demonstrate the suitability of Ga:YIG for magnonic applications, enabling operation of spin waves in various nanostructures while maintaining comparable spin-wave properties across different geometries.

The group velocity of spin waves significantly influences their decay length $L_\mathrm{d}$, a critical parameter that determines the propagation distance achievable within the waveguide. Its measurements were conducted in waveguides with three different widths, and the results are presented in Fig.~\ref{fig:fig3}. The decay length reaches a notable value of \qty{7.43}{\micro \meter} even in the narrowest waveguide with a width of \qty{145}{\nano\meter} (Fig.~\ref{fig:fig3}(a)). Furthermore, $L_\mathrm{d}$ increases in wider waveguides (Fig.~\ref{fig:fig3}(b, c)), demonstrating the robust propagation characteristics of Ga:YIG nanostructures. The slight deviation from the exponential fit can be explained by the presence of non-linear effects in the spin-wave excitation. This phenomenon was previously observed by Heinz et al.~\cite{heinz2020propagation} in non-substituted YIG nanowaveguides. Due to the enhanced group velocities and still relatively low Gilbert damping achieved in Ga:YIG~\cite{bottcher2022fast, dubs2025magnetically}, spin waves propagate here over larger distances in comparison to the experimentally investigated non-substituted YIG nanowaveguides~\cite{heinz2020propagation}.

Additionally, Ga:YIG provides a robust platform for a mode- and wavevector-selective spin-wave excitation. The negative slope in the dispersion relation of non-substituted YIG results in the simultaneous excitation of two spin waves with different wavevectors for some frequencies, one in the dipolar and one in the exchange regime~\cite{klingler2014design}. Being dominated by the exchange interaction, spin waves in Ga:YIG provide a single-wavevector excitation in the entire frequency range of interest. Furthermore, the strong frequency separation between modes in Ga:YIG compared to non-substituted YIG allows for the easier establishment of a single-mode operation regime in a broad frequency range (from \qty{6.45}{\giga \hertz} to \qty{8.14}{\giga \hertz} in the particular case shown in Fig.~\ref{fig:fig2}(a)) even in the wider waveguides, providing access to long distance magnon transport in complex magnonics networks.

\section{Conclusion}

In this work, we have comprehensively investigated the spin-wave transport properties of Ga:YIG nanowaveguides through a combined experimental, numerical, and analytical approach. Our findings demonstrate that Ga:YIG represents a significant advancement in magnonic materials, offering superior performance characteristics compared to both non-substituted YIG and metallic ferromagnetic systems.

The key achievements of this study include the demonstration of high spin-wave group velocities reaching up to \qty{600}{\meter \per \second} in nanowaveguides with widths as narrow as \qty{145}{\nano\meter}. These values represent a several-fold increase over the group velocities typically attainable in non-substituted YIG nanowaveguides, where performance is significantly limited by dipolar effects. The linear scaling of group velocity with wavevector in Ga:YIG, arising from its exchange-dominated dispersion relation, enables operation with fast spin waves across a broad frequency range.

Furthermore, we have shown that Ga:YIG nanowaveguides exhibit exceptional spin-wave propagation characteristics, with decay lengths reaching \qty{7.43}{\micro \meter} (for $k_x = \qty{11.25}{\radian \per \micro \meter}$) even in the narrowest structures and extending up to \qty{10.2}{\micro \meter} in wider waveguides.

The isotropic nature of exchange-dominated spin waves in Ga:YIG provides an additional crucial advantage, as demonstrated by the width-independent group velocities observed across different waveguide geometries. This characteristic enables reliable and predictable spin-wave operation in complex magnonic circuits, overcoming the geometry-dependent limitations encountered in conventional dipolar systems.

The combination of high group velocities, long propagation distances, and isotropic behavior positions Ga:YIG as an suitable platform for advancing integrated magnonics and realizing practical magnonic logic devices at the nanoscale. These results open new possibilities for energy-efficient spin-wave-based computing architectures and contribute significantly to the development of next-generation magnonic technologies.

\section*{Acknowledgments}

This research was funded in whole or in part by the Austrian Science Fund (FWF) 10.55776/P34671, 10.55776/I6068, 10.55776/PAT3864023, 10.55776/ESP526, and 10.55776/PIN1434524. CzechNanoLab project LM2023051 funded by MEYS CR is gratefully acknowledged for the financial support of the measurements/sample fabrication at CEITEC Nano Research Infrastructure. C.D. acknowledges the funding by the Deutsche Forschungsgemeinschaft (DFG, German Research Foundation) within the project 271741898 and thanks O. Surzhenko for the VSM measurements and R. Meyer for the technical support. R.V. acknowledges support by the NAS of Ukraine, project \#0124U000270. Q.W. acknowledges the financial support from the National Natural Science Foundation of China (Grant No. 12574118). M.U. acknowledges project No. CZ.02.01.01/00/22\_008/0004594 (TERAFIT).

\bibliography{bib}

\end{document}